\documentclass[]{jpsj2}
\usepackage{graphicx}
\usepackage{latexsym}
\usepackage{amsmath}
\usepackage{color}

\def\ba{\begin{align}}

\title{%
Two-Dimensional Island Shape Determined by Detachment 
}

\author{%
Yukio \textsc{Saito}\thanks{yukio@rk.phys.keio.ac.jp}
and Ryo \textsc{Kawasaki}
}

\inst{%
Department of Physics, Keio University, Yokohama 223-8522 
}

\recdate{\today}

\abst{%
Effect of an anisotropic detachment on a heteroepitaxial island shape
is studied by means of a kinetic Monte Carlo simulation of a square lattice gas
model.
Only with molecular deposition followed by surface diffusion, 
islands grow in a ramified dendritic shape, similar to DLA.
Introduction of molecular detachment from edges makes islands compact.
To understand an anisotropic island shape 
observed in the experiment of pentacene
growth on a hydrogen-terminated Si(111) vicinal surface,
asymmetry in detachment around the substrate step is assumed.
Edge molecules detach more to the higher terrace than to the lower terrace.
The island edge from which molecules are easy to detach is smooth 
and the one hard to detach is dendritic.
If islands are close to each other, islands tend to
align in a line, since detached molecules from the smooth edge of
the right island are fed to the dendritic and fast growing edge of the
left island.
}

\kword{%
heteroepitaxial growth, island shape anisotropy, detachment, lattice gas model,
kinetic Monte Carlo simulation
}

\begin{document}
\sloppy
\maketitle

\section{Introduction}
For the production of high quality devices, it becomes
more and more important to control atomic structures on clean crystalline surfaces.
There are many studies, for example, on a shape and density
of two-dimensional (2D) islands 
which are nucleated and grown on a substrate surface
by molecular beam epitaxy.
\cite{barabasi+95,pimpinelli+98,venables00,michely+03}
They are mainly determined through the competition 
between the diffusion of adsorbed molecules
on a substrate surface and their incorporation kinetics
at the island edge.
If the molecule incorporation at edges are very fast and
diffusion controls the growth, island edges undergo a morphological 
instability to form dendrites.
In the extreme case of irreversible solidification, 
an island takes a ramified irregular shape with many branches, 
\cite{hwang+91}
similar to a diffusion-limited aggregate (DLA).
\cite{witten+81,witten+83}
Its structure has no characteristic length and is fractal.
On the other hand, 
if the incorporation kinetics is slow and limits the island growth, 
islands becomes compact. 
\cite{bartelt+94,bales+95,saito03}

On a vicinal surface steps exist and may bring about
 additional effects on surface structures.
For a homoepitaxial case, they lead to surface roughening via
various instabilities as meandering and bunching.
\cite{misbah+,bales+90,uwaha+92,saito+94,stoyanov91,sato+95}
For a heteroepitaxial case, substrate steps may affect a shape of 
epilayer islands. 
It is in fact observed recently in
a heteroepitaxial growth of a pentacene (Pn) submonolayer
on a vicinal surface of hydrogen-terminated Si(111) [H-Si(111)] substrate.
\cite{nishikata+07}
The experiment is performed in an expectation that
vicinal steps give control on the epifilm orientation.
Pn wets the substrate surface completely and form 2D
flat islands in a submonolayer coverage. 
When a terrace width is large on a vicinal H-Si(111) surface, Pn
islands are isotropic and compact in shape.
In contrast, when the slope of the vicinal substrate becomes steep and
the step density increases such that an island covers many steps, 
the shape becomes anisotropic;
on the higher side of the vicinal surface an island is compact
and rounded, whereas on the lower side it is dendritic.
The island shape is found to be independent of the impinging orientation 
of the Pn molecular beam.
As for the origin of this shape anisotropy, no
definite conclusion has yet been stated.\cite{nishikata+07}

There is another experiment where anisotropy is observed in the shape
of two-dimensional islands; a thin film of organic molecules on oxidized
Ga-As substrate. \cite{kambe+07}
The substrate has no steps, and the shape anisotropy
is attributed to the tilting of long organic molecules in a crystalline film.
Precise mechanism to determine shape anisotropy for this sytem still awaits 
for an analysis.

Here we propose and study a possibility that 
may induce a shape anisotropy of islands; 
anisotropy in the rate of molecular detachment from an island edge.
For example, on a vicinal surface, 
the rate of detachment from the edge of an island 
that climbs up a substrate step may
differ from that from the edge of a descending island. 
Even on a flat substrate surface, if long molecules are 
adsorbed in a tilted conformation, 
detachment rate of molecules from the edge of an island may
depend on a relative directions of tilting and migration.

Whatever the mechanism may be, an anisotropy in the detachment rate should be
reflected in the island shape.
This is the aim we address ourselves in the present paper.
We consider a simple and thus generic model 
where point molecules are adsorbing on a flat square substrate, and
study it by kinetic Monte Carlo simulations \cite{bortz+75}.
The model is described in \S2.
In order to understand the effect of detachment on the island shape,
we first study the case of an isotropic detachment in \S3.
In \S4 we describe the possible origin of anisotropic detachment
referring to
the Pn experiment by way of example, 
and study the island shape in an anisotropic case.
The last section \S5 summarizes the results.

\section{Deposition and  Diffusion on a Square Substrate}
Our model consists of point molecules depositing on a flat substrate surface 
of a square lattice.
They diffuse on the substrate, and
when they meet each other during the surface diffusion,
 they attach each other to form a dimer. 
When a diffusing molecule touches an already existing
cluster, it attaches the cluster to increases its size, and 
the cluster grows to an island.
When a molecule is deposited above an island, it diffuses on
the island terrace till it reaches an edge, and steps it down
to be incorporated into the island.
Evaporation of adsorbed molecule from the surface is excluded,
since the growth is performed under a low temperature.
With only surface diffusion and attachment at island edges, 
grown islands are ramified \cite{hwang+91}
and look fractal, similar to DLA
\cite{witten+81,witten+83}, at a low coverage.
In order to realize compact islands, detachment or edge diffusion
of molecules weakly bonded to the island edge are necessary unless 
evaporation is allowed.
In the present model we allow detachment of those molecules connected to
the island edge with only one nearest-neighboring bond, for simplicity.
Still, they can be detached only to the substrate surface, but 
not allowed to hop up on the epitaxial island: There are three possible
orientations for them to detach.

We now review theories on an island density and  
shapes in terms of three parameters; 
\cite{pimpinelli+98,venables00,michely+03}
a deposition flux $F$, a surface diffusion constant $D_s$,
and a detachment rate $D_e$ from a singly-connected edge site.
As for the average island distance $\ell_2$ in a stationary state, 
we have to consider two-dimensional nucleation on a flat substrate.
For a freshly deposited molecule, it takes about a time 
$\tau_2=\ell_2^2/D_s$ 
to migrate a region of a linear dimension $\ell_2$.
During this time interval, other molecules are deposited 
to a density $n_1=F \tau_2$, and the initial molecule has a 
probability $n_1 \ell_2^2$ to meet another molecule during its migration.
(Logarithmic correction is necessary for more detailed analysis.)
By assuming that the encounter causes the instantaneous nucleation, the 
encounter probability is the same with the nucleation probability  per site.
On the other hand, the latter is given by $(a/\ell_2)^2$ with $a$ being
the lattice constant,
since there is one nucleus within a region of $\ell_2^2$.
By equating the two results, one obtains
$n_1 \ell_2^2=(a/\ell_2)^2 $ or a scaling relation 
\begin{align}
\ell_2=\Big( \frac{D_sa^2}{F} \Big)^{1/6}.
\label{eq1}
\end{align}
This scaling holds when solidification takes place irreversibly and
the critical size of a nucleation is $i_c=1$.
In a more general case with a critical nucleation size $i_c$,
the island separation is given by\cite{pimpinelli+98,michely+03}
\begin{align}
\ell_2 \sim \Big( \frac{D_s}{F} \Big)^{\frac{i_c}{2(i_c+2)}},
\label{eq2}
\end{align}
 where the lattice constant
is set to unity.

An irreversibly grown island with $i_c=1$
has a ramified profile similar to DLA, and consists of many edge molecules 
loosely connected only by a single bond. These loose molecules may detach from
the island with a rate $D_e=D_s e^{-J/k_BT}$ at a temperature $T$
with an energy penalty $J$ of breaking a bond.
On a square lattice, the detached molecule can migrate three open directions
on the substrate.
If detachment happens along
a straight island edge, it corresponds to an edge diffusion.
If detachment takes place on the tip of a pointed branch, the detached molecule
becomes a freely migrating molecule. But the fast terrace diffusion $D_s$
may drive the molecule to another edge site not too far from the detached
site. Therefore, detachment is effectively regarded similar to the
edge diffusion.

When the detachment is possible, the dendritic branches dissolve
and the island can be compact in a polygonal shape.
There is a theory on a stability limit of an edge length $\ell_1$
at which the polygonal shape becomes unstable.
\cite{bartelt+94,bales+95,michely+03}
On a straight edge of an  island of a length $\ell_1$, there should be a kink
which incorporates molecules diffusing on the edge.
With an edge diffusion constant $D_e$, it takes about a time 
$\tau_1=\ell_1^2/D_e$ before a freshly deposited molecule on an edge is
 incorporated into the kink site.
During this period, there are about $F \ell_2^2 \tau_1$ molecules
deposited within the territory of an island,
 where $\ell_2$ is an average separation between islands.
 With a fast surface diffusion $D_s$, 
this number of molecules are deposited on the island edge.
Thus, when it is of order unity, a one-dimensional
nucleation takes place on the edge and the polygon side would become
rough kinetically. Therefore, the critical size to
keep an island in a polygonal shape is given as 
\begin{align}
\ell_1 \sim \Big( \frac{D_e}{F \ell_2^2} \Big)^{1/2}.
\label{eq3}
\end{align}

\begin{figure}[h]
\begin{center} 
\includegraphics[width=0.32\linewidth]{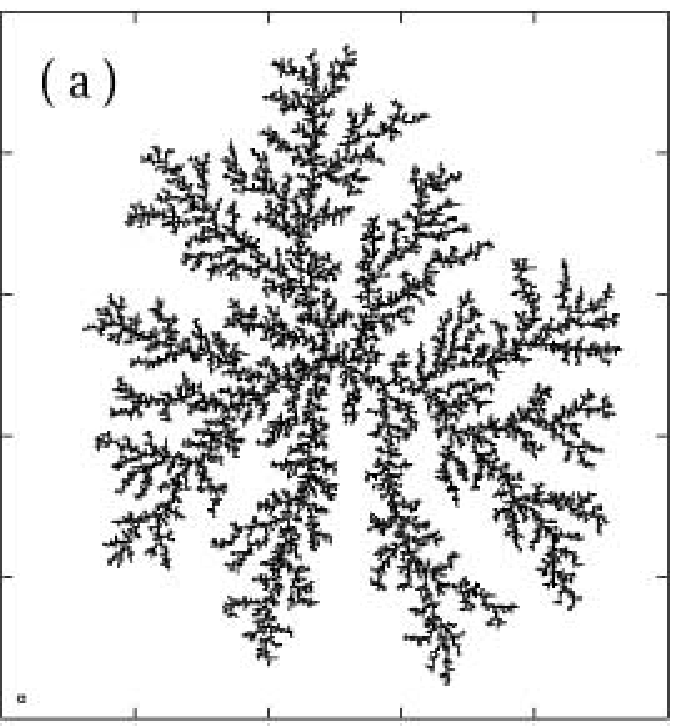}
\includegraphics[width=0.32\linewidth]{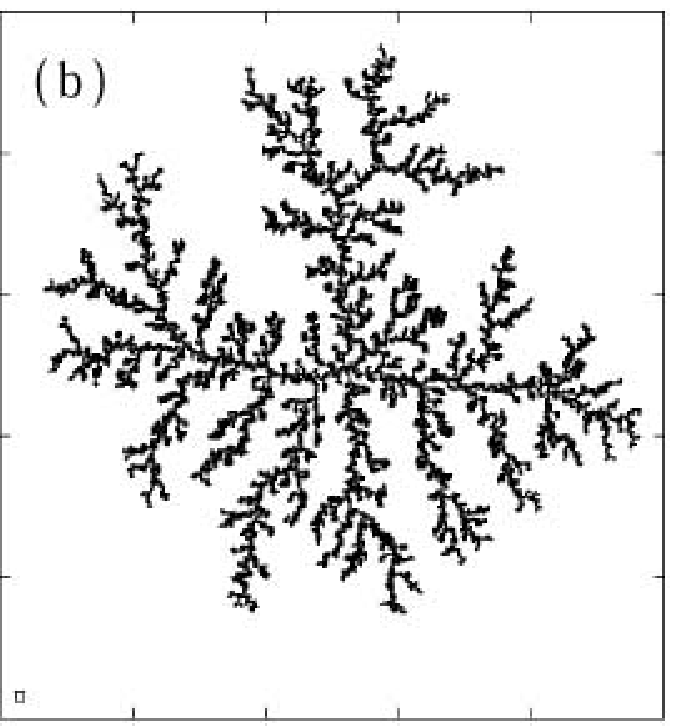}
\includegraphics[width=0.32\linewidth]{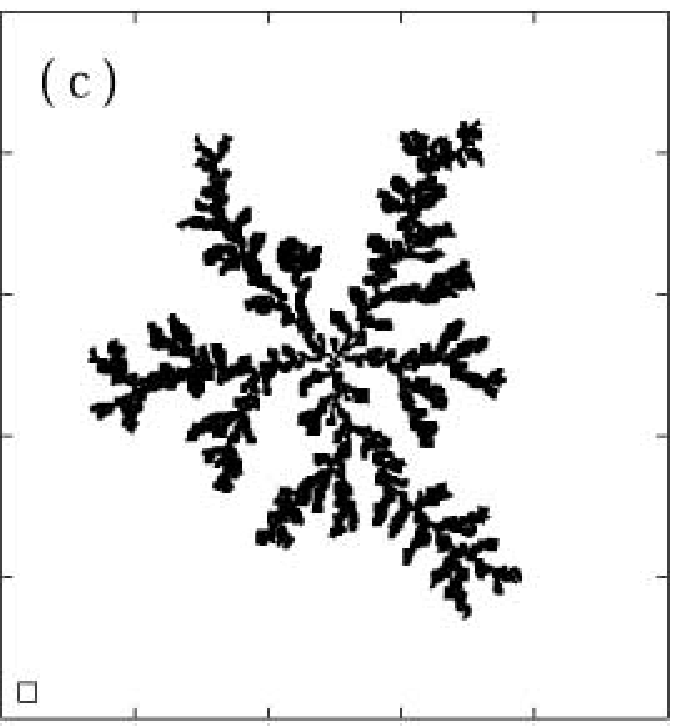}
\includegraphics[width=0.32\linewidth]{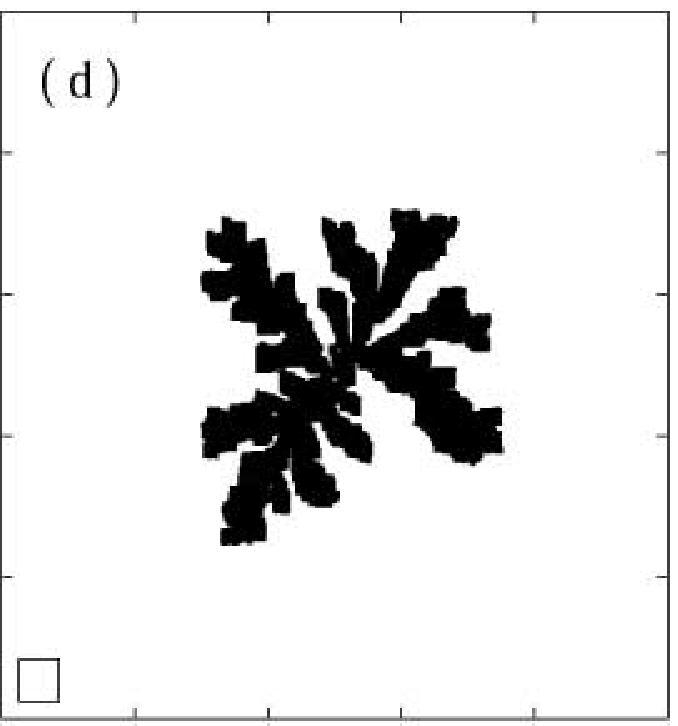}
\includegraphics[width=0.32\linewidth]{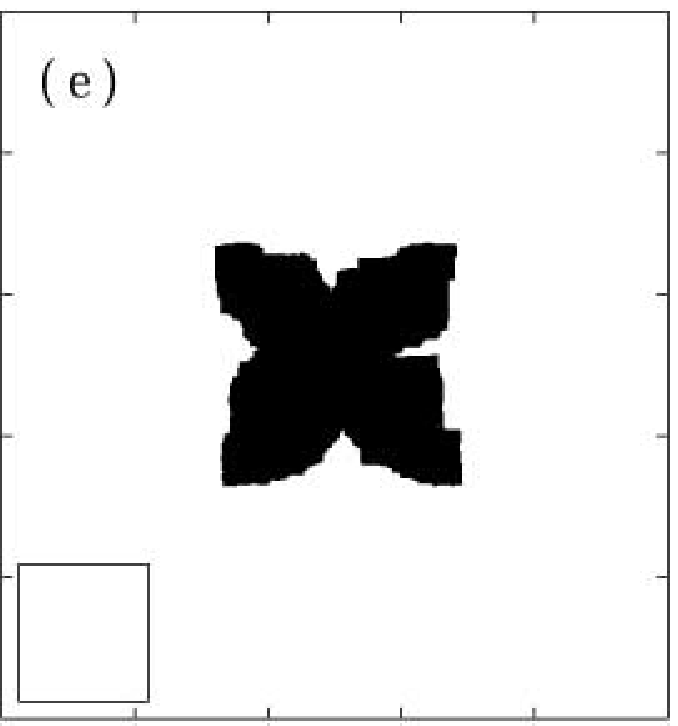}
\includegraphics[width=0.32\linewidth]{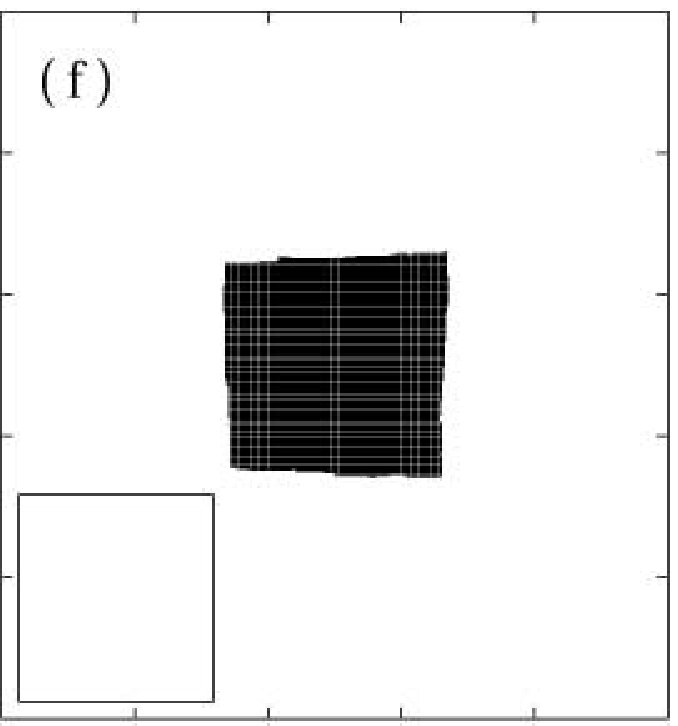}
\end{center} 
\caption{An island shape at the coverage $\Theta=0.1$ on a substrate with
a size $500^2$.
The ratio of the deposition flux $F$ and the surface diffusion constant
$D_s$ is $D_s/F=10^{15}$.
The detachment rate $D_e/F$ is (a) 0, (b) $10^{3}$,
(c) $10^{4}$, (d) $10^{5}$, (e) $10^{6}$, (f) $10^{7}$. 
A square at the left bottom is a maximum one that fits in 
the corresponding aggregate.
}
\label{fig1}
\end{figure}

\section{Isotropic Detachment}

In order to understand the effect of detachment on the island shape,
we perform kinetic Monte Carlo simulations
\cite{bortz+75} of a system with deposition, diffusion and detachment
processes.
Before the study of an anisotropy effect,
we start from the more fundamental case with an isotropic detachment 
in this section.
To concentrate on the island shape variation,
parameter values are so chosen as to allow only a single island in a
large system with a linear dimension of $L=500$. 
In this large system, we provide initially
a nucleus of a size $2 \times 2$ at the center of the system;
This is the minimum size of an immobile island 
under a finite detachment rate, $i_c=3$.
For a DLA-like island without detachment $D_e=0$,
 the size of a critical island is $i_c=1$ and  
the ratio of the surface diffusion $D_s$ to the
deposition flux  $F$ as high as $D_s/F=10^{15}$ predicts a sufficiently large 
island separation as $\ell_2 \sim (D_s/F)^{1/6} \sim 300$. 
A single island is, in fact, realized as shown in Fig.1(a)
at the coverage of $\Theta=0.1$.
The dimension $d_f$ of the aggregate is estimated 
by assuming a scaling relation between the aggregate density $c(r)$ 
and the distance $r$ from the mass center as
\begin{align}
c(r) \propto r^{d_f-2}.
\end{align}
If the aggregate is compact, the density is independent of the distance $r$
and one obtains $d_f=2$. For a fractal DLA, its dimension is well-known 
to be $d_f=1.7$, less than a space dimension 2. 
\cite{witten+81,barabasi+95}
Our simulation data is consistent with this value, though
a deviation is observed around the center which is exposed for a long time
under deposition;  spaces between
dendritic branches will be filled under a perpetual deposition, and
the island becomes compact eventually.

We now introduce detachment of molecules from the island edge 
to the substrate surface.
The grown island at a coverage $\Theta=$0.1 becomes fat and anisotropic,
as shown in Figs. 1(b) to (f).
Every island consists of $N_a=25,000$ molecules.
As $D_e/F$ increases from $10^{3}$ to $10^5$, shown 
in Fig. 1(b) to (d), 
each branch becomes fat.
On further increase of detachment rate to $D_e/F=10^{6}$ , the island
becomes skeletal with a corner instability 
due to the Berg effect, as shown in Fig. 1(e).
At a further large value of $D_e/F=10^7$ (Fig. 1(f)) and more, 
the island remains square up to a coverage $\Theta=$0.1.

On looking at island shapes in Fig. 1 
one notices that there is a certain length scale
which characterizes the width of dendritic branches,
and this scale increases as the detachment rate $D_e$ increases.
To quantify this characteristic length, one may think of various ways.
The density $c(r)$ is expected to show crossover from the central compact 
region to the outer fractal region. However, the central region is susceptible
to small size effect and large fluctuation, and it is difficult to
define a small crossover length definitely by this method.
Another means is the box counting method used to characterize fractal objects.
One covers the object by squares of a linear size $b$ and 
counts the minimum number $N(b)$ of boxes to completely cover up the object.
It is proportional to $b^{-2}$ for a compact
object and to $b^{-d_f}$ for a fractal.
If the object is compact up to a length scale $\ell_1$, then the number
$N(b)$ is expected to show a crossover behavior. 
In the present case, an island at $D_e=0$ shows a fractal behavior
and that at $D_e/F=10^7$ a compact one, but for intermediate values
of $D_e$ the crossover turns out rather smooth and it is difficult to estimate
the crossover length uniquely.
Still another possibility is to investigate the edge length defined as the number
of molecules with one to three nearest neighbor molecules.
It is proportional to the island size $N_a$ 
for a fractal object and to $\sqrt{N_a}$ for a compact object. 
As the island size increases,
one expects a crossover from compact to fractal.
In fact, on the contrary, for small $D_e/F \le 10^{5}$, 
the perimeter length first increases in proportion to $N_a$
and then deviates from the linear relation. 
Because of the fast surface diffusion, islands first ramify, 
and then afterwards the profile is healed by a slow detachment process.

\begin{figure}[h]
\begin{center} 
\includegraphics[width=0.49\linewidth]{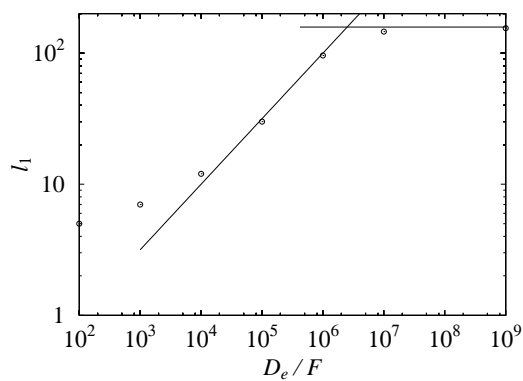}
\end{center} 
\caption{Characteristic length $\ell_1$ of an island versus
the detachment rate $D_e$.
A slope represents the scaling behavior $\ell_1 \propto D_e^{1/2}$. 
}
\label{fig2}
\end{figure}

After all the incompetence, we estimate the characteristic length as
a linear size of a maximum square box which can be completely covered by
the island.
Square boxes placed at the left bottom corner in Fig. 1 have the
characteristic linear size $\ell_1$ thus determined.
Due to the symmetry, there may be two kinks on an edge of a square
island, and the length thus defined may corresponds to twice the
characteristic length defined in eq. (\ref{eq3}).
In fact, we observe few kinks within the range of $\ell_1$ on edges.
But these kinks are transient, and we prefer the present way 
to estimate $\ell_1$, since the main body of the island
accumulates all the history of the growth and averages out fluctuation.
For the scaling law (\ref{eq3}), in any case, a numerical factor is irrelevant,
and we plot the length $\ell_1$ 
as a function of the detachment rate $D_e$
in Fig.2.

At the coverage $\Theta=$0.1 in a system of a size $L^2=500^2$,
the island has a size $N_a=25,000$, and the maximum linear size
of a square island is $\sqrt{N_a} \approx 158$, which is indicated by a horizontal
line in Fig.2.
For large $D_e/F \ge 10^7$, the size $\ell_1$ reaches this maximum limit.
In the intermediate range of $10^{4} \le D_e/F \le 10^{6}$,
the scaling behavior 
\cite{bartelt+94} $\ell_1 \propto D_e^{1/2}$ of eq. (\ref{eq3})
is found compatible with our data, as shown in Fig. 2.
At small $D_e$, on the other hand,
deviation from the scaling law eq. (\ref{eq3}) is apparent.
This is related to our definition of $\ell_1$.
At small $D_e$ an island looks similar to DLA,
and the outer dedritic branches in Figs. 1(a) and (b)
are a very fine.
However, the
central portion of an island is exposed to a constant deposition flux
all through the time until the coverage reaches to $\Theta=0.1$.
Therefore, the dendrite may thicken in the center. 
In contrast, for the DLA diffusing molecules start always
far away from the aggregate itself, and central region is screened
from the deposition flux.
Thus, the central portion of the present island can cover-up 
large squares which are not covered in the outer region 
or in the true DLA.
Since our estimation of $\ell_1$ requires only a single square box
which is completely covered by an island,
the present method overestimates $\ell_1$,
especially for small $\ell_1$ at small $D_e$.
This causes the deviation from the scaling law at small $D_e$ in Fig. 2.

\section{Anisotropic Detachment}

We now consider a shape anisotropy induced by an anisotropic
detachment rate. The problem may be relevant to the shape
anisotropy observed in the heteroepitaxial growth of Pn islands
on a vicinal surface of H-Si(111) substrate \cite{nishikata+07}
or of an organic epifilm on a flat oxidized GaAs substrate. \cite{kambe+07}
In the following a Pn system is chosen by way of an example 
to explain the possible origin of a detachment anisotropy, 
but the applicability of our generic model is
by no means restricted to this specific system.

Pn is a flat, elongated molecule, and on H-Si(111) surface
it is known to grow a wetting film with the molecular long axis perpendicular
to the surface, namely in a standing-up orientation
\cite{shimada+05}. 
It means that Pn interaction with the substrate surface is weaker than the
Pn-Pn interaction.
On a flat terrace, deposited Pn molecules make surface diffusion
and they are incorporated into a 2D island when they encounter.
Island are compact and their shape is isotropic.
If the incorporation were instantaneous and irreversible,
the island should be ramified and fractal, in contradiction to the experiment.
A detachment from the island edge may accomplish an observed compact 
island shape on a flat terrace, as discussed in the previous section.
We here study an asymmetry  observed in island shape when Pn grows
on a vicinal H-Si(111) surface. As the step  density in the vicinal surface
increases
and an island spans over many steps, island edges at the higher level are round
whereas lower edges become dendritic.

Steps may affect molecular diffusion on a substrate surface.
If the step enhances an energy barrier against surface diffusion, 
the effect is symmetric around the step. The rate of step-up diffusion
slows down by the same amount as that of step-down diffusion.
Since no net flow is expected in this case, islands should be symmetric.
If a step edge makes an extra bonding with a diffusing molecule, 
it will reside preferentially in front of the step.
Then, nucleation rate will be enhanced in front of the step, 
and consequently islands will be nucleated along it,
as often observed in various experiments.
In the Pn experiment,\cite{nishikata+07} however, this kind of 
island alignment along steps is not observed.
Therefore, these kind of step effects on surface diffusion may be negligible
for the Pn experiment.

Instead, we consider a step effect on molecular 
detachment from island edges.
Imagine that a Pn crystalline film grows in a step-down direction and
reaches an upper side of the descending step. Then,
further crystal growth takes place by incorporating
a Pn molecule from the lower terrace. In this case, the whole length of a
newly attached molecule interacts with
the island edge and the step ledge, and it is hard to be detached back to the
lower terrace.
The detachment rate is diminished by a factor $R_L<1$.
On the contrary, when the film grows in the step-up direction
and reaches the lower side of the ascending step,
the next Pn molecule to be incorporated is located on the higher terrace,
and its upper part does not contribute to the Pn-Pn bonding. 
Since the substrate step has a height of about one fourth of the length of 
a Pn molecule, 
 a loss in the bonding energy may be significant.
Also, the free upper part of the long molecule 
may be susceptible to thermal fluctuation, and there may be a substantial
entropy contribution which reduces the bonding free energy.
Accordingly, those molecules incorporated into an island edge
from the higher terrace are easily detached back with
an enhancement factor $R_H>1$.
We now study the effect of this anisotropic detachment with $R_H$ and $R_L$
on the island morphology.

\begin{figure}[h]
\begin{center} 
\includegraphics[width=0.32\linewidth]{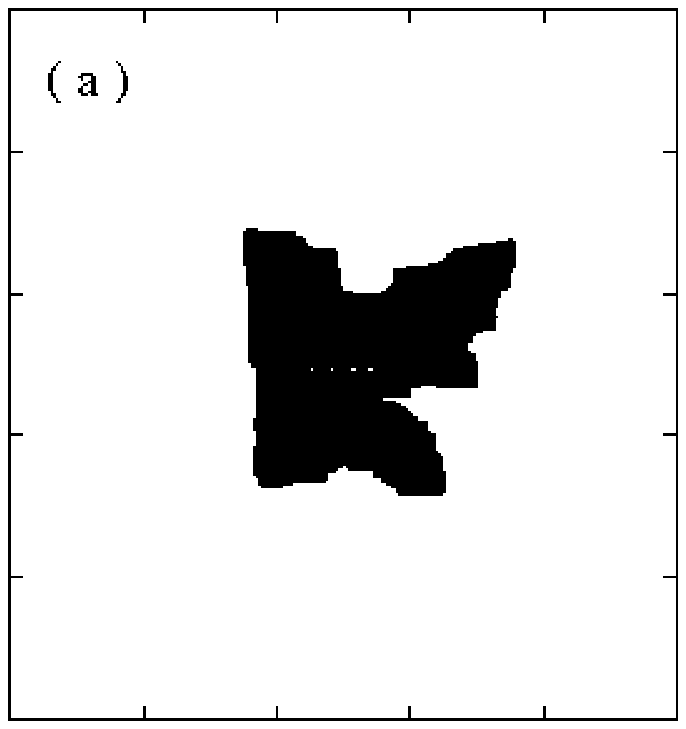}
\includegraphics[width=0.32\linewidth]{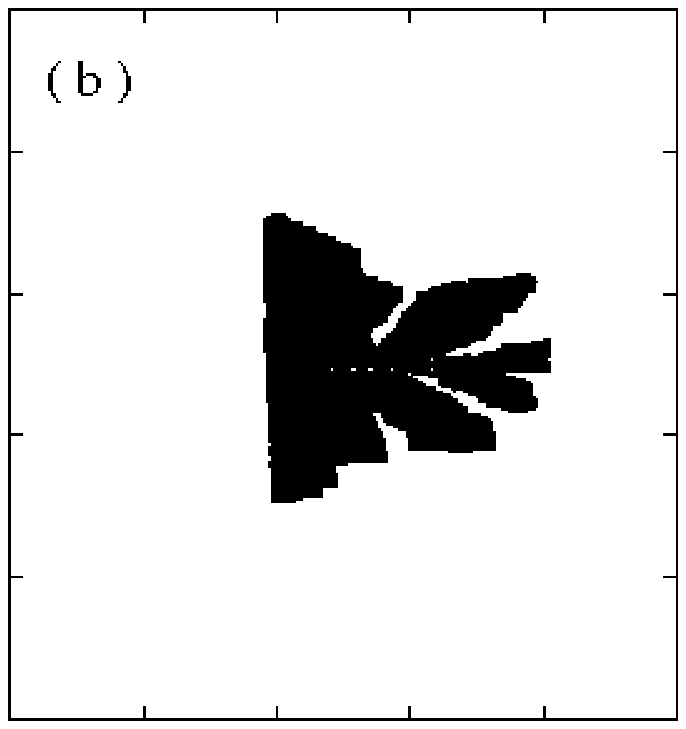}
\includegraphics[width=0.32\linewidth]{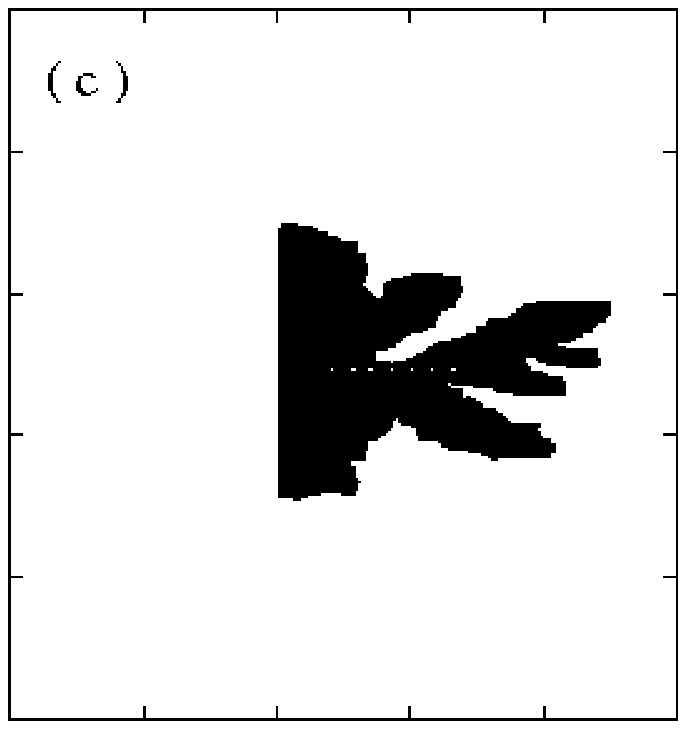}
\end{center} 
\caption{Island shapes at different detachment rate from the left and the right edges. 
The surface diffusion is $D_s/F=10^{15}$ and the detachment rate is $D_e/F=10^6$.
Detachment to the higher terrace (to the left) is enhanced  
by a factor $R_H$ and 
to the lower terrace (to the right) is reduced by $R_L$,
 and their values are ($R_H, R_L$)=
(a) (2, 0.5), (b) (5, 0.2) and (c) (10, 0.1).
The coverage is $\Theta=0.1$ on a substrate with a size $500^2$.
}
\label{fig3}
\end{figure}

As we want to find out the essential factors of shape determination,
the model here is chosen to be as simple as possible:
Instead of providing a vicinal surface tilted down in $+x$ direction, 
we simulate a model of heteroepitaxial growth on a flat substrate surface,
 but with anisotropic detachment rate.
The molecule that resides on the left edge of an island 
and connected to the island only by one bond has 
a higher detachment rate $D_e R_H$ with $R_H > 1$, whereas
that on the right edge detaches from the island less frequently with 
the rate $D_e R_L$ and $R_L < 1$. 
Those molecules residing on the upper and the lower edge of the island
detach as before with the rate $D_e$.
The diffusion constants are set to $D_s/F=10^{15},~D_e/F=10^{6}$
to allow a single island in our system
of a size $L^2=500^2$.
By initially providing an embryo of a size 2 by 2 at the center of the system,
an island is nucleated and grows to
a shape as shown in Fig. 3 at the coverage $\Theta=0.1$.
The detachment enhancement factor to the higher terrace $R_H$ 
is chosen to be inversely proportional to the suppression factor
 to the lower terrace $R_L$
 in order to decrease the number of free parameters, and
they are varied as
(a) ($R_H,~R_L$)=(2, 1/2), (b) 
(5, 1/5), 
and (c) (10, 1/10).
In the isotropic case with $R_H=R_L=1$, the island has a skeletal shape
shown in Fig.1(e). Even though enhancement and the suppression factors,
$R_H$ and $R_L$, 
are applicable everywhere in the system and there is no local inhomogeneity, 
the left and the right halves of the island show a drastic contrast:
The left-hand side (lhs) is polygonal and 
the right-hand side (rhs) is dendritic.

To quantify the shape difference in the lhs and the rhs of an island,
one may think to utilize fractal dimensions in both sides. 
In Fig. 3(c), for instance, the lhs of an island is very narrow, 
and a box counting in a limited width
yields only a rough evaluation as
$d_f \approx 1.9 \sim 2$.
On the rhs, 
dendrites are clearly observable but they are still fat and not long enough.
A fractal dimension depends on the narrow fitting range
as $d_f \approx 1.7 \sim 1.8$.
For a more accurate determination of $d_f$
larger systems are necessary, but it is out of scope of the
present study.

As for the upper and lower edges in Fig. 3, 
the lhs and the rhs of an island also show a contrast.
In fact, the detachment rates from them are the same $D_e$ as in 
Fig. 1 (e).
The lhs of islands in Fig. 3 are similar to that in Fig. 1(e), whereas
the upper and lower edges in the rhs of an island in Fig. 3(c)
are quite flat compared to the rhs of an island in Fig. 1(e).
It may be because branching of the upper or lower edges to the left 
( or to $-x$ direction) is suppressed. 
Branching takes place only to the right, 
and these dendrites grow to the right ( or $+x$ direction ) fast. 
This suppression of left branching decreases fluctuation 
of upper and lower edges of an island.

Furthermore, on looking through Fig. 3(a) to (c), one may get an impression that
an island seems to be shifted to the right
as $R_L$ decreases.
This apparent shift is due to the difference in the growth rate
in $\pm x$-directions. The dendritic right edge
grows faster than the facetted left edge.
Difference in the two edge speeds
is quantified in a previous experiment \cite{nishikata+07}
by a ratio of growth distances of side edges from the
nucleation center.
Since the initial nucleus is provided at the center of the system 
$x_c=y_c=L/2$,
the degree of the shape anisotropy is measured by the maximum and the minimum 
distance in the $x$ direction of the island edges 
from the initial nucleation center as
\begin{align}
\frac{\mathrm{L}}{\mathrm{S}}=\frac{x_{max}-x_c}{x_c-x_{min}}.
\label{eq5}
\end{align}
The simulation result is summarized in Fig. 4.
As the reduction factor $R_L$ of detachment to the lower terrace decreases,
the shape anisotropy increases. 

\begin{figure}[h]
\begin{center} 
\includegraphics[width=0.5\linewidth]{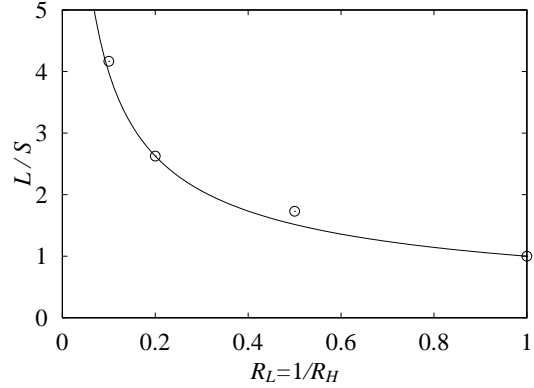}
\end{center} 
\caption{Island shape anisotropy $L/S$ versus detachment anisotropy $R_L=1/R_H$
obtained from Fig. 3. Line is a guide for eyes.
The surface diffusion is $D_s/F=10^{15}$ and the detachment rate is $D_e/F=10^6$.
}
\label{fig4}
\end{figure}
So far, we discussed the shape of a single island.
In a large area of the vicinal surface many islands are nucleated 
and they affect their growth and shape each other through the diffusion field.
Their average separation, for instance, is determined by the ratio
of the surface diffusion constant $D_s$ and the deposition flux $F$,
if detachment anisotropy is discarded, 
as described in \S 2.
With anisotropic detachment, we obtained spatial distribution of
islands as shown in Fig. 5.
The surface diffusion constant in Fig. 5(a) and 
(c) 
is set to 
$D_s/F=10^{10}$ and in 
Fig. 5(b)
 to $D_s/F=10^{9}$.
Detachment rate in Fig. 5(a) is set $D_e/F=10^4$ with the enhancement 
factor $R_H=10$ and reduction factor $D_L=0.1$.
Islands are well separated and they seem to be distributed homogeneously.
In Fig. 5(b), the surface diffusion is decreased to $D_s/F=10^9$, and
the island density increases. Then one may detect some correlation in 
the alignment of islands
in $x$-direction.
In Fig. 5(c) the surface diffusion is kept the same with the case
in Fig. 5(a) but, instead, the detachment rate at the upper and the lower edges
of an island is increased by a factor of ten:
Molecules are easily detached from the three edges except the right one,
and accordingly the right dendritic edge grows faster. 
Enhanced molecule detachment in $y$ direction enlarges the longitudinal
island separation and weakens the correlation in this direction.
In contrast, 
one has an impression that 
islands align laterally in $x$ direction.

In order to quantify structural correlation, 
structure function or the Fourier transform of the
density correlation is usually employed.
In the present small system, however, the structure function is noisy
and it is difficult to identify the anisotropy of the island alignment.
Instead, we use a simple measure to indicate an anisotropy 
in $x$ and $y$ directions; 
number densities of empty rows and columns 
which are free of adsorbed molecules.
For example, in Fig. 5(a) 4.8\% of the vertical columns are free of adsorbed molecules whereas 9.2\% of the lateral rows are free.
Though each island has a clear anisotropic shape, islands arrangement in space
is rather isotropic.
In Fig. 5(b), the empty row increases to the density of 16.6\%,
 whereas there are no empty columns; 
there are at lease one adsorbed molecule on every column. 
In Fig. 5(c) alignment in $x$ direction tends stronger;
51\% of empty rows but no empty columns.
Thus, the measure of empty rows and columns seems very clear and helpful to
indicate the degree of anisotropic arrangement of islands.

\begin{figure}[h]
\begin{center} 
\includegraphics[width=0.32\linewidth]{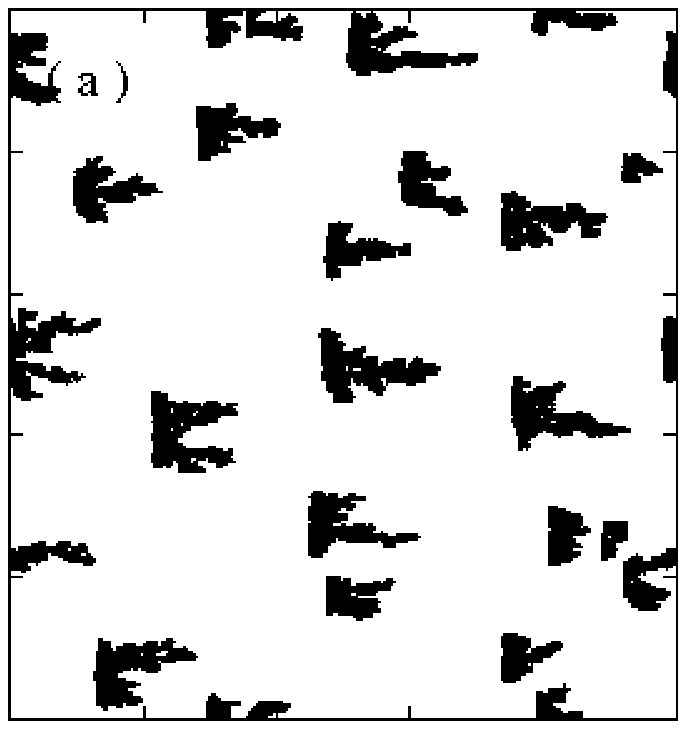}
\includegraphics[width=0.32\linewidth]{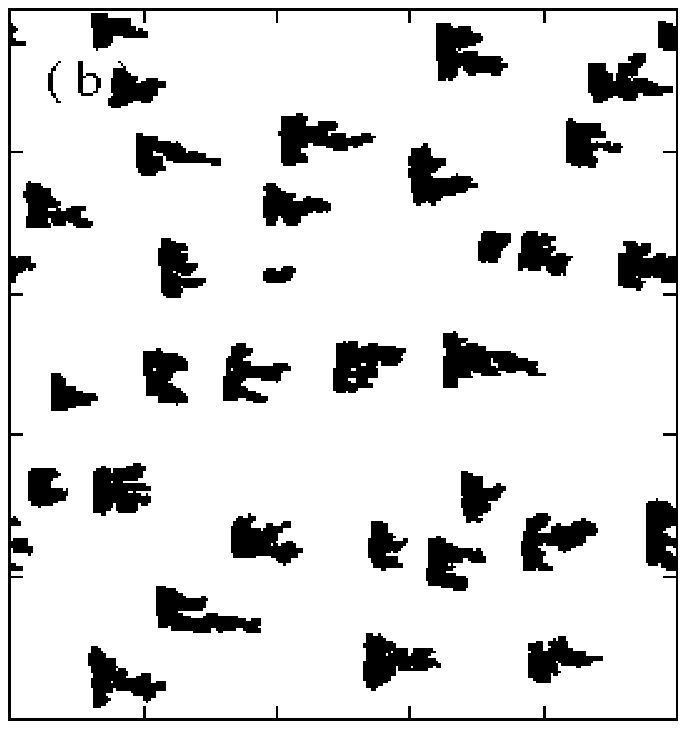}
\includegraphics[width=0.32\linewidth]{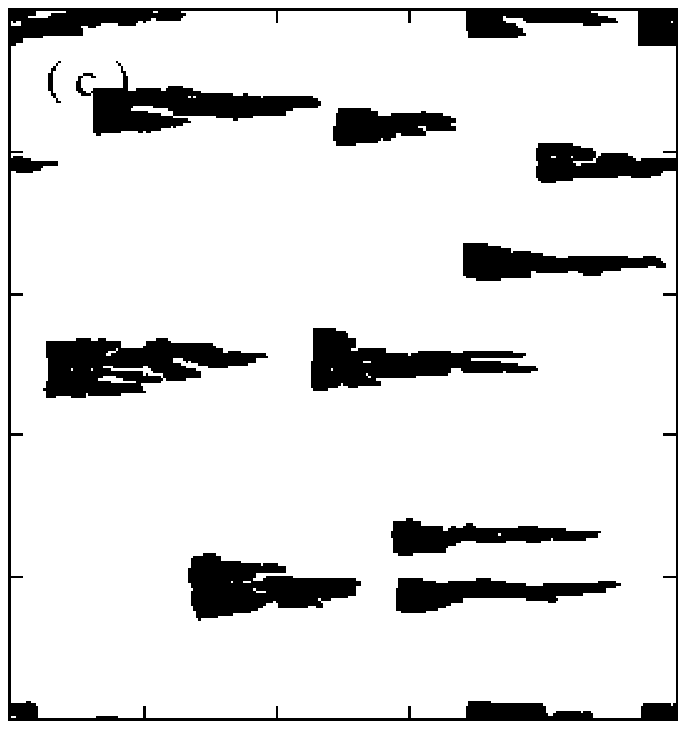}
\end{center} 
\caption{Islands distribution at small surface diffusion,
(a) and (c) $D_s/F=10^{10}$, and (b) $D_s/F=10^{9}$.
Other parameters are
(a) and  (b) $ D_e/F=10^4,~R_H=10, R_L=0.1$,
and
(c) $ D_e/F=10^5,~R_H=1, R_L=0.01$.
The system size is $500^2$ and the coverage is $\Theta=0.1$.
}
\label{fig5}
\end{figure}

\section{Summary and Discussions}

An effect of an anisotropic detachment from island edges
 on the determination of a two-dimensional island shape
is studied by means of a kinetic Monte Carlo simulation of a square lattice gas
model of heteroepitaxial growth.
The study may be relevant to the formation of anisotropic islands observed 
in heteroepitaxial growth experiments of Pn molecules on a vicinal surface
\cite{nishikata+07}
and/or of tilted organic molecules on a flat substrate.
\cite{kambe+07}
Only with a molecular deposition with flux $F$ 
followed by a surface diffusion $D_s$, 
islands take a ramified dendritic shape, similar to DLA.
Addition of detachment from the edge with a rate $D_e$ 
lets dendritic branches fat, 
and with a large $D_e$ compact square islands are formed.

By assuming that a detachment rate depends on an edge orientation,
island shape becomes anisotropic.
For example, the left edge with an enhanced detachment with a rate
$R_H D_e$ with $R_H>1$ is found smooth and compact, whereas
the right edge with a small detachment rate $R_L D_e$ with $R_L<1$ 
is ramified and dendritic.
If islands are close to each other, one notices a tendency that islands
align linearly, since detached molecules from the smooth edge of
the right island are fed to the dendritic and fast growing edge of the
left island.

We comment on the anisotropy factors, $R_H$ and $R_L$, in Pn experiments.
\cite{nishikata+07} 
Even without steps, Pn islands have compact shape, and thus
the detachment process takes place often enough.
This means that a Pn-Pn bonding is not too strong to prevent 
edge molecules from detachment, and some part of the bonding is broken
thermally. Since the Pn molecule is long,
it is conceivable that the upper part is most susceptible to 
thermal fluctuation.
When substrate steps are present, a Pn molecule incorporated from the 
higher terrace misses bonding at its upper part, but if thermal fluctuation
already breaks the bonding there, the effect will be small. 
Then, $R_H$ can be close to unity.
For a Pn molecule attaching from the lower terrace, on the other hand, 
its upper part may make a strong bond with the pre-existing island edge 
which lies higher.
Then, the detachment may be strongly supressed, yielding a very small $R_L$.
Estimation of precise values of these factors requires microscopic 
information of chemical bondings, molecular rigidities, and so on,
and out of scope of the present study.

Above explanation is refering to the Pn epilayer on a vicinal surface
 just by way of example,
but it is not meant to be specific to that system.
We expect our model is universal and is applicable to other cases, such
as heteroepitaxial growth of tilted molecules on a flat substrate surface.
\cite{kambe+07}

\section*{Acknowledgment}
We acknowledge G. Sazaki and S. Nishikata, and T. Nakada and T. Kambe
for showing their experimental data prior to the publication. 
Part of the work is supported by a Grant in Aid for Scientific Research 
from Japan Society of the Promotion of Science. 
Y. S. is benefited from the interuniversity 
cooperative research program of the Institute for Materials Research, 
Tohoku University.


\end{document}